\documentclass[aps,prl,preprint,tightenlines,superscriptaddress,showpacs,byrevtex]{revtex4}

\usepackage{epsfig}% Include figure files
\newcommand{\mb}{{M_{\rm bc}}}
\newcommand{\de}{{\Delta{E}}}
\newcommand{\plpi}{{p\bar{\Lambda}\pi^-}}

\newcommand{\llk}{{\Lambda\bar{\Lambda}K^+}}
\newcommand{\llpi}{{\Lambda\bar{\Lambda}\pi^+}}
\newcommand{\mll}{{M_{\Lambda \bar{\Lambda}}}}

\newcommand{\llkbronly}{2.91 ^{+0.90}_{-0.70}}
\newcommand{\llkbr}{(2.91 ^{+0.90}_{-0.70} \pm 0.38) \times 10^{-6}}

\newcommand{\llkbra}{{(2.91\,^{+\,0.90}_{-\,0.70} \; ({\rm stat})\pm 0.38 \; ({\rm syst})) \times 10^{-6}}}

\newcommand{\llpiul}{< 2.8 \times 10^{-6}}
\newcommand{\llb}{\Lambda\bar{\Lambda}}

\begin{document}

%\doubles

%% >>>> For summer conference papers, add this \preprint command

%\preprint{\tighten\vbox{\hbox{\hfil Belle-Note-569}
%                        \hbox{\hfil Parallel Session: 8 }
%                        \hbox{\hfil ABS726 }
%                        \hbox{\hfil hep-ex/nnnn, if available}
%}}

%\twocolumn[\hsize\textwidth\columnwidth\hsize\csname
%@twocolumnfalse\endcsname

\epsfysize 25mm \epsfbox{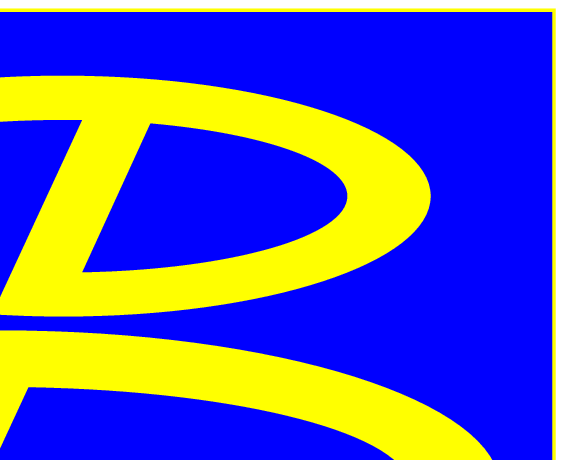}
\begin{flushright}
\vskip -25mm \noindent \hspace*{3.0in}{\bf Belle Prerpint 2004-17
\\KEK Prerpint 2004-24}
\end{flushright}
%\vskip 25mm

\title{ \quad\\[1cm] \Large
Observation of $B^+ \to \llk$}

% to make it single spaced
\tighten
%%% Paper:    Observation of B+ -> Lambda Lambda-bar K+
%%% Journal:  Physical Review Letters
%%% Contacts: Y.-J. Lee (listeve@hep1.phys.ntu.edu.tw)
%%%           M.-Z. Wang (mwang@hep1.phys.ntu.edu.tw)
%%% Non-responding authors or those who said NO are commented out.
%%% ====================================================================
%%% Click the RELOAD button on your web browser to see the updated file.
%%% ====================================================================
%%% Use \input{author} to insert this material into your latex file.
%%%%% Force institutions to appear in alphabetical order when typeset.
%%%\affiliation{Aomori University, Aomori}
\affiliation{Budker Institute of Nuclear Physics, Novosibirsk}
\affiliation{Chiba University, Chiba} \affiliation{Chonnam
National University, Kwangju}
%%%\affiliation{Chuo University, Tokyo}
\affiliation{University of Cincinnati, Cincinnati, Ohio 45221}
%%%\affiliation{University of Frankfurt, Frankfurt}
\affiliation{Gyeongsang National University, Chinju}
\affiliation{University of Hawaii, Honolulu, Hawaii 96822}
\affiliation{High Energy Accelerator Research Organization (KEK),
Tsukuba} \affiliation{Hiroshima Institute of Technology,
Hiroshima} \affiliation{Institute of High Energy Physics, Chinese
Academy of Sciences, Beijing} \affiliation{Institute of High
Energy Physics, Vienna} \affiliation{Institute for Theoretical and
Experimental Physics, Moscow} \affiliation{J. Stefan Institute,
Ljubljana} \affiliation{Kanagawa University, Yokohama}
\affiliation{Korea University, Seoul}
%%%\affiliation{Kyoto University, Kyoto}
\affiliation{Kyungpook National University, Taegu}
\affiliation{Swiss Federal Institute of Technology of Lausanne,
EPFL, Lausanne} \affiliation{University of Ljubljana, Ljubljana}
\affiliation{University of Maribor, Maribor}
\affiliation{University of Melbourne, Victoria}
\affiliation{Nagoya University, Nagoya} \affiliation{Nara Women's
University, Nara} \affiliation{National Kaohsiung Normal
University, Kaohsiung} \affiliation{National United University,
Miao Li} \affiliation{Department of Physics, National Taiwan
University, Taipei} \affiliation{H. Niewodniczanski Institute of
Nuclear Physics, Krakow} \affiliation{Nihon Dental College,
Niigata} \affiliation{Niigata University, Niigata}
\affiliation{Osaka City University, Osaka} \affiliation{Osaka
University, Osaka} \affiliation{Panjab University, Chandigarh}
\affiliation{Peking University, Beijing} \affiliation{Princeton
University, Princeton, New Jersey 08545}
%%%\affiliation{RIKEN BNL Research Center, Upton, New York 11973}
%%%\affiliation{Saga University, Saga}
\affiliation{University of Science and Technology of China, Hefei}
\affiliation{Seoul National University, Seoul}
\affiliation{Sungkyunkwan University, Suwon}
\affiliation{University of Sydney, Sydney NSW} \affiliation{Tata
Institute of Fundamental Research, Bombay} \affiliation{Toho
University, Funabashi} \affiliation{Tohoku Gakuin University,
Tagajo} \affiliation{Tohoku University, Sendai}
\affiliation{Department of Physics, University of Tokyo, Tokyo}
\affiliation{Tokyo Institute of Technology, Tokyo}
\affiliation{Tokyo Metropolitan University, Tokyo}
\affiliation{Tokyo University of Agriculture and Technology,
Tokyo}
%%%\affiliation{Toyama National College of Maritime Technology, Toyama}
\affiliation{University of Tsukuba, Tsukuba}
%%%\affiliation{Utkal University, Bhubaneswer}
\affiliation{Virginia Polytechnic Institute and State University,
Blacksburg, Virginia 24061} \affiliation{Yonsei University, Seoul}
  \author{Y.-J.~Lee}\affiliation{Department of Physics, National Taiwan University, Taipei} % Taiwan
  \author{M.-Z.~Wang}\affiliation{Department of Physics, National Taiwan University, Taipei} % Taiwan

  \author{K.~Abe}\affiliation{High Energy Accelerator Research Organization (KEK), Tsukuba} % KEK
  \author{K.~Abe}\affiliation{Tohoku Gakuin University, Tagajo} % TohokuGakuin
% \author{N.~Abe}\affiliation{Tokyo Institute of Technology, Tokyo} % TIT
  \author{T.~Abe}\affiliation{High Energy Accelerator Research Organization (KEK), Tsukuba} % KEK
% \author{I.~Adachi}\affiliation{High Energy Accelerator Research Organization (KEK), Tsukuba} % KEK
  \author{H.~Aihara}\affiliation{Department of Physics, University of Tokyo, Tokyo} % Tokyo
% \author{M.~Akatsu}\affiliation{Nagoya University, Nagoya} % Nagoya
% \author{M.~Asai}\affiliation{Hiroshima Institute of Technology, Hiroshima} % Hiroshima
  \author{Y.~Asano}\affiliation{University of Tsukuba, Tsukuba} % Tsukuba
% \author{T.~Aso}\affiliation{Toyama National College of Maritime Technology, Toyama} % Toyama
  \author{V.~Aulchenko}\affiliation{Budker Institute of Nuclear Physics, Novosibirsk} % BINP
  \author{T.~Aushev}\affiliation{Institute for Theoretical and Experimental Physics, Moscow} % ITEP
  \author{T.~Aziz}\affiliation{Tata Institute of Fundamental Research, Bombay} % Tata
% ...
% Ljubljana
% \author{D.~Z\"urcher}\affiliation{Swiss Federal Institute of Technology of Lausanne, EPFL, Lausanne}
\collaboration{The Belle Collaboration}

\normalsize

\begin{abstract}

We report the first observation of the charmless hyperonic $B$
decay, $B^+ \to \llk$, using a 140 fb$^{-1}$  data sample recorded
at the $\Upsilon({\rm 4S})$ resonance with the Belle detector at
the KEKB $e^+e^-$ collider. The measured branching fraction is
${\mathcal B}(B^+ \to \llk) = \llkbr$. We also perform a search
for the related decay mode $B^+ \to \llpi$, but do not find a
significant signal. We set a 90\% confidence-level upper limit of
${\mathcal B}(B^+ \to \llpi) \llpiul$.

% \vskip1pc \pacs{PACS
%numbers: 13.20.H }

\end{abstract}

\maketitle

{\renewcommand{\thefootnote}{\fnsymbol{footnote}}

\setcounter{footnote}{0}
%\newpage

\normalsize

%
%

%\onecolumn

Charmless hadronic $B$ decays are of great interest since they
provide opportunities for probing CP violation, as well as for
testing our understanding of strong interactions.
While charmless mesonic modes were first established over ten
years ago \cite{Battle:1993rw}, $B$ decays to charmless baryonic
final states such as $p\bar pK^+$ \cite{ppk}, $p\bar \Lambda
\pi^-$ \cite{plpi}, $p\bar pK^0$, $p\bar p\pi^+$ and $p\bar
pK^{*+}$ \cite{Wang:2003iz} were first seen only recently. The
branching ratios for these three-body decays are larger than those
for two-body modes such as $B^0\to p\bar p$, for which only upper
limits have been reported \cite{2body}. Another intriguing feature
is the threshold peaking behavior commonly observed in the baryon
pair mass spectrum \cite{ppk,plpi,Wang:2003iz}. Both features were
anticipated by theory \cite{HS}, but the underlying dynamics are
still far from understood
\cite{PP,CY,CHT,suzuki,newpara,Rosner:2003bm}.

In this paper we report the observation of $B^+ \to \Lambda\bar
\Lambda K^+$ decay, the first example of a charmless $B$ decay to
a final state containing two hyperons \cite{conjugate}.
The rate is found to be comparable to that of other charmless
three-body baryonic modes as well as that for $B\to \phi\phi K$
\cite{phiphik}. The invariant mass of the $\Lambda \bar{\Lambda}$
system has a prominent near-threshold peak.

With three strange particles in the final state, the $\llk$ mode
may complement $b\to s\bar ss$ dominated mesonic modes such as
$B\to \phi K^{(*)}$
\cite{kfjack:2003jf,Abe:2003yt,Aubert:2003mm,Aubert:2004ii}, for
which the polarization and CP asymmetry may be sensitive to new
physics. With the three-body final state and self-analyzed
polarization information from the $\Lambda$ decay
\cite{HS,CHT,suzuki,newpara}, the $B^+ \to \Lambda\bar \Lambda
K^+$ process can be used to probe not only CP violation, but T
(time reversal symmetry) violation as well.

 We use a data sample of 140 fb$^{-1}$ integrated luminosity, consisting
of $152$ million $B\overline{B}$ pairs with no accompanying
particles,
collected by the Belle detector %on the $\Upsilon({\rm 4S})$ resonance
at the KEKB asymmetric energy $e^+e^-$ (3.5 on 8~GeV)
collider\cite{Kurokawa:2001nw}. The Belle detector is a large
solid angle magnetic spectrometer that consists of a three-layer
silicon vertex detector (SVD), a 50-layer central drift chamber
(CDC), an array of aerogel threshold \v{C}erenkov counters (ACC),
a barrel-like arrangement of time-of-flight scintillation counters
(TOF), and an electromagnetic calorimeter comprised of CsI(Tl)
crystals located inside a super-conducting solenoid coil that
provides a 1.5~T magnetic field.  An iron flux-return located
outside of the coil is instrumented to detect $K_L^0$ mesons and
to identify muons. The detector is described in detail
elsewhere~\cite{Belle}.

Since KEKB operates with a center-of-mass energy at the
$\Upsilon({\rm 4S})$ resonance, which decays into a
$B\overline{B}$ pair, one can use the following two kinematic
variables to identify the reconstructed $B$ meson candidates: the
beam constrained mass, $\mb = \sqrt{E^2_{\rm beam}-p^2_B}$, and
the energy difference, $\de = E_B - E_{\rm beam}$, where $E_{\rm
beam}$, $p_B$, and $E_B$ are the beam energy, the momentum, and
energy of the reconstructed $B$ meson in the $\Upsilon({\rm 4S})$
rest frame, respectively. The $\mb$ resolution of about 3
MeV/$c^2$ is dominated by the beam energy spread. The $\de$
resolution for $B^+\to\llk$ ranges from 12 MeV to 17 MeV,
depending on $\mll$.

The event selection criteria are based on the information obtained
from the tracking system (SVD+CDC) and the hadron identification
system (CDC+ACC+TOF), and are optimized using Monte Carlo (MC)
simulated event samples.

All primary charged tracks are required to satisfy track quality
criteria based on the track impact parameters relative to the
interaction point (IP). %, which is determined run-by-run.
The deviations from the IP position are required to be within
$\pm$0.3 cm in the transverse ($x$-$y$) plane, and within $\pm$3
cm in the $z$ direction, where the $z$ axis is the opposite of the
positron beam direction.

Primary kaon and pion candidates are selected based on $K/\pi$
likelihood functions obtained from the hadron identification
system. To identify kaons/pions, we require the likelihood ratio
$L_{K(\pi)}/(L_K+L_\pi)$ to be greater than 0.6. For kaons(pions),
this requirement has an efficiency of 86\%(89\%) and a pion(kaon)
misidentification probability of 8\%(10\%). $\Lambda$ candidates
are reconstructed via the $p\pi^-$ decay channel using the method
described in ref.~\cite{2body}.

The dominant background for the rare decay modes reported here is
from $e^+e^- \to q\bar{q}$ continuum processes (where $q = u, d,
s, c$). The background from generic $B$ decays and known baryonic
$B$ decays is negligible. This is confirmed using an off-resonance
data set (10 fb$^{-1}$) taken 60 MeV below the $\Upsilon({\rm
4S})$ and MC samples of generic $B$ decay, 150 million continuum
events and known baryonic $B$ decays. In the $\Upsilon({\rm 4S})$
rest frame, continuum events tend to be jet-like while
$B\overline{B}$ events tend to be spherical. We follow the scheme
defined in~\cite{etapk} that combines 7 shape variables to form a
Fisher discriminant~\cite{fisher} in order to optimize continuum
background suppression. The variables used have almost no
correlation with $\mb$ and $\de$. Probability density functions
(PDFs) for the Fisher discriminant and the cosine of the angle
between the $B$ flight direction and the beam direction in the
$\Upsilon({\rm 4S})$ frame are combined to form the signal
(background) likelihood ${\cal L}_{\rm s (b)}$. We require the
likelihood ratio ${\cal R} = {\cal L}_{\rm s}/({\cal L}_{\rm
s}+{\cal L}_{\rm b})$ to be greater than 0.4; this suppresses
about 73\% of the background while retaining 88\% of the signal.
The selection point is determined by optimizing $S/\sqrt{S+B}$,
where $s$ and $b$ denote the number of signal and background; here
a signal branching fraction of $4\times10^{-6}$ is assumed.
We also require only one candidate per event. In the case of
multiple $B$ candidates (about 2.6\% of the events), we choose the
candidate with the highest ${\cal R}$ value.
The signal PDFs are determined using the signal MC simulation; the
background PDFs are obtained from the data sideband events with
$5.2$ GeV/$c^2<\mb<$ 5.26 GeV/$c^2$ and 0.1 GeV$<|\de|<$ 0.3 GeV.

\begin{figure}[t!]
\centering \epsfig{file=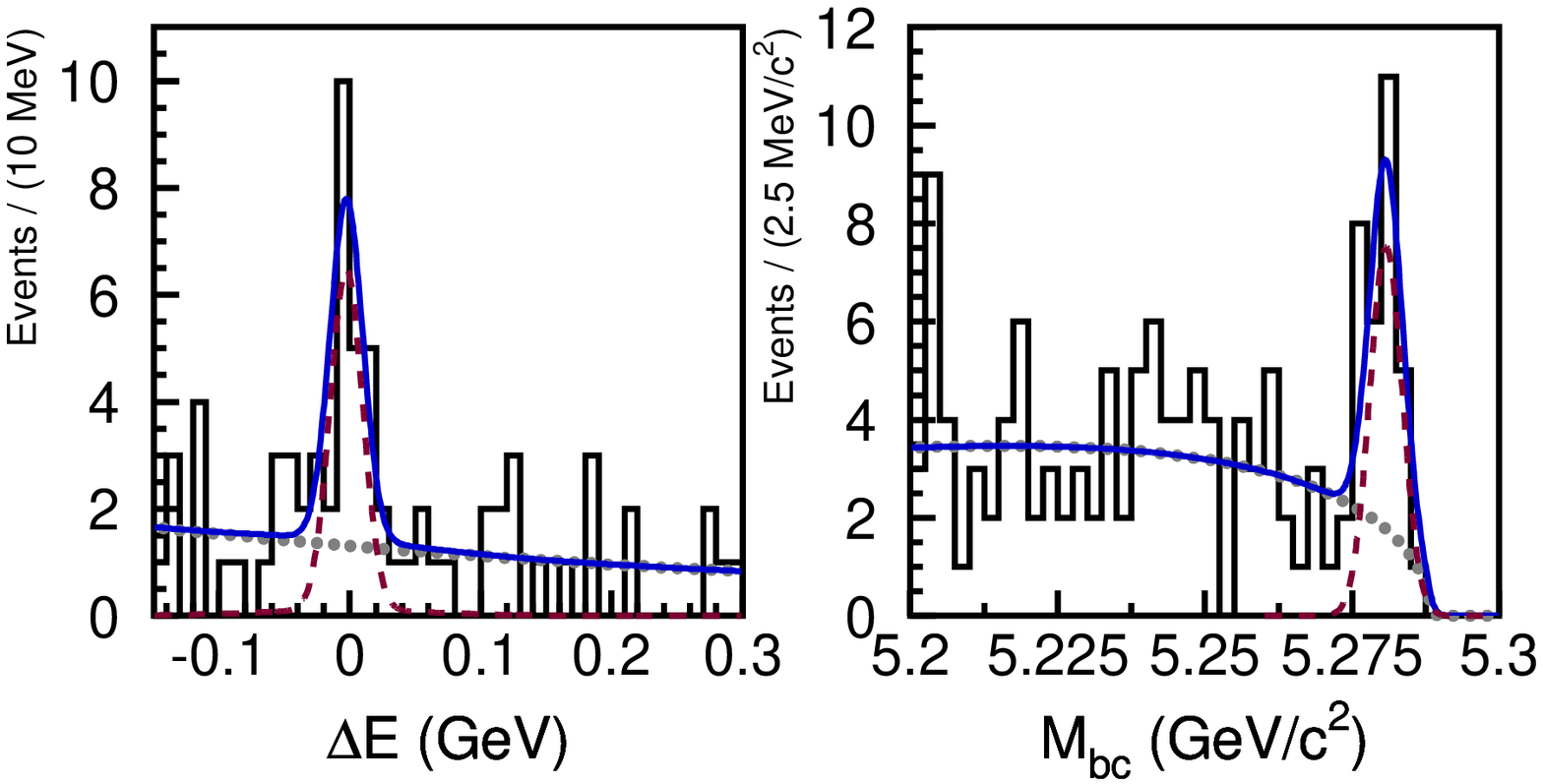,width=3.3in} \centering
\caption{$\de$ and $\mb$ distributions of $B^+ \to \llk$
candidates for $\mll<2.85$ GeV/$c^2$. The solid and dashed curves
represent the fit results and the signal, respectively; the dotted
curve shows the background contribution.} \label{fg:llkmbde}
\end{figure}

To ensure the decay is charmless, we exclude 2.85
 GeV/$c^2<\mll<$ 3.128 GeV/$c^2$ and 3.315 GeV/$c^2<\mll<$ 3.735
GeV/$c^2$ regions to remove contributions from $J/\psi$, $\eta_c$,
$\psi^\prime$ and $\chi_{c0,1,2}$ in order to extract the three
body decay branching fraction.

We perform an unbinned extended maximum likelihood fit to the
events with $-0.15$ GeV$<\de<$ 0.3 GeV and $\mb>$ 5.2 GeV/$c^2$ to
estimate signal yields. The extended likelihood function $\cal L$
is
\[\small
 {\cal L} =e^{-(N_s+N_b)} \prod_{i=1}^{N}
[N_sP_s(M_{{\rm bc}_i},\Delta{E}_i)+ N_bP_b(M_{{\rm
bc}_i},\Delta{E}_i)],
\]
where $P_s(P_b)$ is the signal(background) PDF and $N_s(N_b)$
denotes the number of signal(background) candidates. The signal
PDF is the product of a Gaussian function, which represents $\mb$,
and a double Gaussian for $\de$. The means and the widths of the
signal PDFs are determined by MC simulation. Differences between
data and MC are corrected by the $B^-\to D^0\pi^-$ and $D^0\to
K^-\pi^+\pi^-\pi^+$ control sample.

We use the parametrization first suggested by the ARGUS
collaboration~\cite{Argus}, $ f(\mb)\propto \mb\sqrt{1-(\mb/E_{\rm
beam})^2} \exp[-\xi (1-(\mb/E_{\rm beam})^2)]$, to model the
background $\mb$ distribution and a 2nd order polynomial for the
background $\de$ shape. We perform a two-dimensional unbinned fit
to the $\de$-$\mb$ distribution, floating the signal and
background normalizations as well as the background shape
parameters.

The $\mb$ distribution (with $|\de|<$ 0.05 GeV) and the $\de$
distribution (with $\mb >$ 5.27 GeV/$c^2$) for the region $\mll<$
2.85 GeV/$c^2$ ({\it i.e.} below charmonium threshold)
 are shown in Fig.~\ref{fg:llkmbde} with fit
results overlaid. The two-dimensional unbinned fit gives a signal
yield of $22.9^{+5.8}_{-4.8}$ with a statistical significance of
$7.4$ standard deviations. The significance is defined as
$\sqrt{-2{\rm ln}(L_0/L_{\rm max})}$, where $L_0$ and $L_{\rm
max}$ are the likelihood values returned by the fits with signal
yield fixed at zero and floating, respectively~\cite{PDG}.

\begin{figure}[b!]
\mbox{\psfig{figure=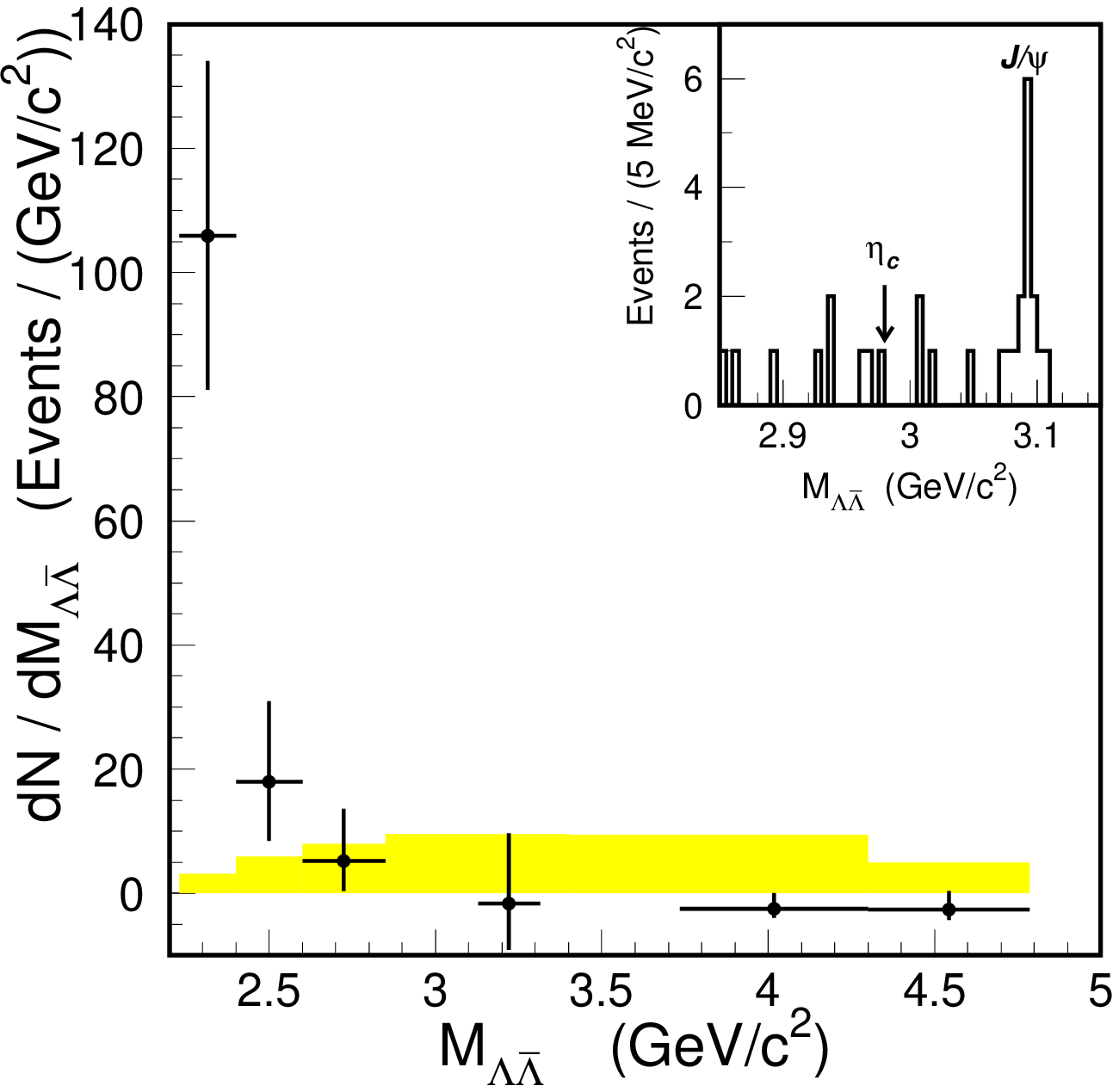,width=3.3in,height=2.6in}}
\caption{Fitted yield divided by bin size for $B^+\to \llk$ as a
function of $\mll$. The shaded distribution is from a phase space
MC simulation with the area normalized to the signal yield. Note
that the charmonium veto has been applied. The inset shows the
$\llb$ mass spectrum for the $\eta_c$ and $J/\psi$ signal regions.
} \label{fg:phase}
\end{figure}

We fit the signal yield in bins of $\mll$ and the result as a
function of $\Lambda\bar{\Lambda}$ mass is shown in
Fig.~\ref{fg:phase}. The observed mass distribution peaks at low
$\Lambda\bar{\Lambda}$ mass, similar to those observed in
\cite{ppk,plpi,Wang:2003iz}. Since the decay is not uniform in
phase space, we calculate the partial branching fraction for each
$\mll$ bin with the corresponding detection efficiency determined
from a large phase space MC sample and an additional special MC
sample with a $\mll$ peak near threshold\cite{thresholdMC}. The
$\Lambda\bar{\Lambda}$ invariant mass spectrum for the events in
the $B^+\to\llk$ signal region ($|\de|<$ 0.05 GeV and $\mb
>$ 5.27 GeV/$c^2$) with $2.85$ GeV/$c^2<\mll<$ 3.15 GeV/$c^2$ is
shown in the inset of Fig.~\ref{fg:phase}. A clear $J/\psi$ signal
is evident.
% We calculate the branching fractions for each bin by
%bin-dependent efficiency.
The results of the fits along with the efficiencies and the
partial branching fractions are given in Table~\ref{bins}. We sum
the partial branching fractions in Table~\ref{bins} to obtain
${\cal B}(B^+ \to \llk)=\llkbra,$ with a statistical significance
of 5.1 standard deviations.

\begin{table}[t!]
\large \caption{Results of the $\de-\mb$ fit, detection
efficiencies ($\epsilon$), and branching fractions (${\cal B}$)
with statistical errors in bins of $\mll$ after the charmonium
veto has been applied. The fit allows the yields to fluctuate
negative. Note that the yields are consistent with zero above
charmonium threshold.} \label{bins}\small
\begin{center}
\begin{tabular}{cccc}
\hline\hline $\mll$(GeV)  & ~~Signal Yield~~& ~~Efficiency(\%)~~&
${\cal B}$ ($10^{-6}$)
\\
%\hline $<2.4$& $16.7^{+4.7}_{-4.0}$& ??.?& $?.??^{+?.??}_{-?.??}$
%\\
\hline $<2.4$& $18.0^{+4.8}_{-4.2}$ & 4.83& $2.45^{+0.65}_{-0.57}$
\\
\hline $2.4-2.6$& $3.6^{+2.6}_{-1.9}$& 4.30&
$0.55^{+0.40}_{-0.29}$
\\
\hline $2.6-2.85$& $1.3^{+2.1}_{-1.2}$& 4.04&
$0.21^{+0.34}_{-0.20}$
\\
\hline $3.128-3.315$& $-0.3^{+2.1}_{-1.4}$& 5.19&
$-0.04^{+0.27}_{-0.18}$
\\
\hline $3.735-4.3$& $-1.4^{+1.4}_{-0.8}$& 6.60&
$-0.14^{+0.14}_{-0.08}$
\\
\hline $>4.3$& $-1.3^{+1.5}_{-0.8}$& 6.90& $-0.12^{+0.14}_{-0.08}$
\\
\hline Total & $19.9^{+6.5}_{-5.1}$& - & $\llkbronly$
\\
 \hline\hline
\end{tabular}
\end{center}
\end{table}

The systematic uncertainty in particle selection is studied using
high statistics control samples.
%Proton identification is
%studied with a  $\Lambda \to p \pi^-$ sample.
Kaon/pion identification is studied with a $D^{*+} \to D^0\pi^+$,
 $D^0 \to K^-\pi^+$ sample.
%the $\cal R$ selection is studied
%with a
%$B^0 \to D^- \pi^+$, $D^- \to K_S \pi^-$ sample
The tracking efficiency is studied with a $D^*$ sample, using both
full and partial reconstruction. Based on these studies, we assign
a 7.8\% error for the tracking efficiency and 0.6\% for kaon/pion
identification.

For $\Lambda$ reconstruction we have an additional error on the
efficiency for off-IP tracks reconstruction, determined from the
difference of $\Lambda$ proper time distributions for data and MC
simulation. For the four tracks from the $\llb$ pair this error
amounts to 6.1\%. By studying the $\Lambda\to p \pi^-$ sample we
assign an error of 1\% for each identified proton. There is also a
1\% error for each $\Lambda$ mass selection and a $0.7\%$ error
for each $\Lambda$ vertex selection~\cite{2body}. Summing the
correlated errors for $\Lambda$ and $\bar{\Lambda}$
reconstruction, we obtain a systematic error of 6.9\% for both
$\Lambda$'s.

Continuum suppression is studied with the topologically similar
$B^-\to D^0\pi^-$, $D^0\to K^-\pi^+\pi^+\pi^-$ sample. By changing
the selection criteria on $\cal R$ in the interval 0 -- 0.4, the
efficiencies of data and MC differ by 3\%.

The systematic uncertainty from fitting is 4.9\%, which is
 studied by varying the parameters of
the signal and background PDFs by $\pm 1\sigma$. The MC
statistical uncertainty and modelling with six $\mll$ bins
contributes a 5.0\% error(obtained by changing the $\mll$ bin
size). The error on the number
of total $B\overline{B}$ pairs is determined to be 0.7\%. %~\cite{Bellenote}
The error from the sub-decay branching fraction of $\Lambda \to
p\pi^-$ is 0.8\%~\cite{PDG}.

We sum the correlated errors linearly and then combine the result
with the uncorrelated ones in quadrature. The total systematic
error is 13.0\%.

We perform a cross-check of this analysis by using the
charmonium-veto events. We measure ${\mathcal B}(B^+\to J/\psi
K^+)$ by following the same analysis procedure with $3.06$
GeV/$c^2<\mll<3.14$ GeV/$c^2$. The signal yield is
11.4$^{+3.9}_{-3.2}$ candidates with a statistical significance of
$7.3$ standard deviations. The obtained product branching fraction
is ${\cal B}(B^+ \to J/\psi K^+)\times{\cal
B}(J/\psi\to\Lambda\bar{\Lambda}) =
(1.55\,^{+\,0.53}_{-\,0.44})\;\times 10^{-6}.$
 By using
${\mathcal B}(B^+\to J/\psi K^+)=(1.01\pm 0.05)\times
10^{-3}$\cite{PDG}, our measured branching fraction is
%\{eqnarray*}
${\cal B}(J/\psi\to\Lambda\bar{\Lambda}) = (
 1.54\,^{+\,0.53}_{-\,0.43}\pm0.20\pm0.08 )\;\times 10^{-3},$
%\end{eqnarray*}
 which is in agreement with the world
average value\cite{PDG}. The third error comes from the
uncertainty of ${\cal B}(B^+ \to J/\psi K^+)$.

We also search for the decay mode
$B^+\to\Lambda\bar{\Lambda}\pi^+$,
 which is an example of a $b\to u\bar{u}d$ process with $s\bar{s}$
 popping. The background from $B^+\to\llk$ is negligible. We perform
 a two-dimensional unbinned likelihood fit to the $\de$-$\mb$
 distribution using the same analysis procedure. No
significant signal is found. The $\mb$ and $\de$ distributions
with fit projections are shown in Fig.\ref{fg:llpi}. We use the
fit results to estimate the expected background, and compare this
with the observed number of events in the signal region in order
to set an upper limit on the yield at the 90\% confidence
level~\cite{Gary}. The estimated background is $37.5 \pm 1.0$, the
number of observed events is 41, the systematic uncertainty is
15\%, and the upper limit yield is 21.7. The efficiency, estimated
from the phase space MC, is found to be 5.05\%. The 90\%
confidence level upper limit for the branching fraction is
${\mathcal B}(B^+ \to \llpi) \llpiul$.
\begin{figure}[t!]
\centering \epsfig{file=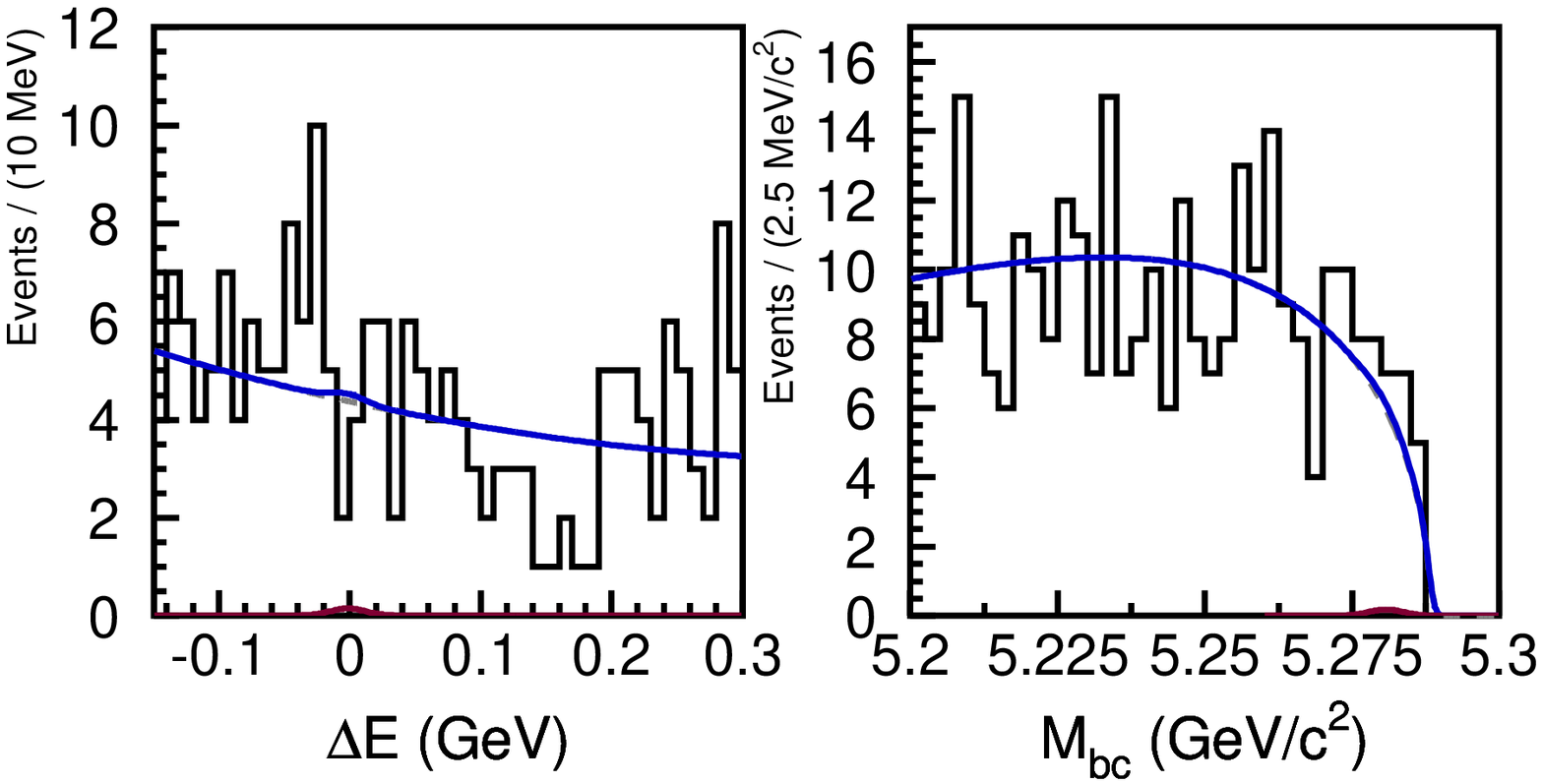,width=3.3in} \centering
\caption{$\de$ and $\mb$ distributions for $B^+ \to \llpi$
candidates. The solid curve is the fit result. } \label{fg:llpi}
\end{figure}

In summary, we have performed a search for the rare baryonic
decays $B^+ \to \llk$ and $\llpi$ with $152$ million $B\bar{B}$
events. A clear signal is seen in the $B^+\to\llk$ mode, where we
measure a branching fraction of ${\mathcal B}(B^+ \to \llk) =
\llkbr$, which is comparable to $B^+ \to p\bar{p}K^+$ and $B^0\to
\plpi$. The observed $\mll$ spectrum peaks toward the threshold as
in the above mentioned modes. This measurement is the first
observation of a $B$ meson decay to a hyperon pair through a $b\to
s \bar{s} s$ process. The $B^+\to\llpi$ mode is not statistically
significant, and we set the 90\% confidence level upper limit
${\mathcal B}(B^+ \to \llpi) \llpiul$.

%\section*{Acknowledgments}
% Please paste this acknowledgement into your latex file.
%***** Acknowledgments *****
We thank the KEKB group for the excellent operation of the
accelerator, the KEK Cryogenics group for the efficient operation
of the solenoid, and the KEK computer group and the NII for
valuable computing and Super-SINET network support.  We
acknowledge support from MEXT and JSPS (Japan); ARC and DEST
(Australia); NSFC (contract No.~10175071, China); DST (India); the
BK21 program of MOEHRD and the CHEP SRC program of KOSEF (Korea);
KBN (contract No.~2P03B 01324, Poland); MIST (Russia); MESS
(Slovenia); NSC and MOE (Taiwan); and DOE (USA).

%\onecolumn

\begin{thebibliography}{99}
\bibitem{conjugate} {Throughout this report, the
inclusion of charge conjugate mode is always implied unless
otherwise stated.}
\bibitem{thresholdMC}
{ We generated a special $B^+\to\llk$ MC sample with 2.23
GeV/$c^2<\mll<$2.85 GeV/$c^2$. The efficiency was obtained using
$1.6\times10^5$ events generated according to phase space and
$5\times10^4$ events of special MC.}
\end{thebibliography}
\end{document}